\documentclass[pra]{revtex4}
\usepackage{amsmath,amssymb,mathrsfs}
\usepackage{psfrag}
\usepackage{graphicx}
\usepackage{graphics}
\usepackage{epsfig}
\usepackage{bm}
\usepackage{color,comment}
\usepackage{verbatim,color,ulem}

\newcommand{\beq}{\begin{equation}}
\newcommand{\eeq}{\end{equation}}
\newcommand{\beqa}{\begin{eqnarray}}
\newcommand{\eeqa}{\end{eqnarray}}
\newcommand{\ba}{\begin{aligned}[b]}
\newcommand{\ea}{\end{aligned}}

\begin{document}

\title{Acoustic plasmons in graphene sandwiched between 
two metallic slabs}

\author{L. Salasnich}
\address{Dipartimento di Fisica e Astronomia ``Galileo Galilei'' 
dell'Universit\`a di Padova, CNISM, and INFN, 
Via Marzolo 8, 35131 Padova, Italy 
\\ 
Padua Quantum Technologies Research Center, 
Universit\`a di Padova, via Gradenigo 6/b, 35131 Padova, Italy
\\
Istituto Nazionale di Ottica (INO) del Consiglio Nazionale 
delle Ricerche (CNR), Via Nello Carrara 1, 50019 Sesto Fiorentino, Italy}

\begin{abstract}
We study the effect of two metallic slabs on the collective dynamics 
of electrons in graphene positioned between the two slabs. 
We show that if the slabs are perfect 
conductors the plasmons of graphene display a linear 
dispersion relation. The velocity of these acoustic plasmons 
crucially depends on the distance between the two metal gates and 
the graphene sheet. In the case of generic slabs, 
the dispersion relation of graphene plasmons is much more complicated 
but we find that acoustic plasmons can still be obtained under 
specific conditions.
\end{abstract}

\maketitle

\section{Introduction} 

In 2004 Graphene, a single layer of carbon atoms arranged in a two-dimensional 
honeycomb lattice, was isolated and characterized \cite{graphene2004}. 
Since then, many electrical, thermal, chemical, optical, and mechanical 
properties of graphene have been studied both experimentally 
and theoretically \cite{graphene2007,graphene2009,graphene2016,graphene2020}. 
Quite remarkably, under appropriate conditions, the 
electrons in graphene, behave as viscous fluids, exhibiting 
peculiar hydrodynamic effects \cite{polini-pt}. In particular, 
it has been shown that the plasmons of graphene display a linear 
dispersion relation when in the proximity of the graphene there 
is a metallic slab which screens the Coulomb potential of electrons 
in graphene \cite{principi,review}. 

In this brief communication we extend the predictions obtained 
in Ref. \cite{principi} by considering a graphene sheet 
sandwiched between two metallic slabs. We find that also in this case 
the electrons of graphene are characterized by acustic modes 
whose dispersion relation is linear in the long wavelength regime. 
We obtain a simple analytical formula for the speed of these acustic modes. 

\section{Graphene sandwiched between two materials} 

The monolayer graphene is a honeycomb lattice of carbon atoms in two 
spatial dimensions. Quasiparticles in graphene have the 
dispersion relation 
\beq 
E_{\bf k}=\pm v_F \hbar |{\bf k}| - \mu \; , 
\eeq
where $v_F$ is the Fermi velocity, 
${\bf k}$ is the two-dimensional (2D) quasiparticle wavevector, 
and $\mu$ the chemical potential. 
The Fermi wavenumber $k_F$ depends on the chemical potential $\mu$ 
through the relation $k_F=\mu/(\hbar v_F)$. Note that in 2D, 
$k_F = \sqrt{4\pi n/g}$, with $n$ the electron number density 
and $g$ the degeneracy. In graphene, $g = 4$: $2$ for spin and $2$ 
for inequivalent valleys 
in the Brillouin zone, and the chemical potential $\mu=\hbar v_F k_F$ is 
usually $\mu \simeq 10^2$ meV, while the Fermi velocity is 
$v_F \simeq 10^6$ m/s \cite{graphene2007,graphene2009,peres}.  

We initially assume that the graphene is sandwiched between two slabs made of 
generic materials, where $L$ is the distance between the two slabs 
and $d$ the distance between the lower slab and the graphene sheet. 
The Coulomb potential of charges in graphene is 
influenced by the the two slabs. 
We choose the $z$ axis perpendicular to the graphene sheet and such 
that $z=0$ fixes the position of graphene sheet. It follows 
that the lower slab is located at $z=-d$ and the upper slab at $z=L-d$.

Within the Random Phase Approximation (RPA) \cite{fetter}, 
the relative dielectric function of graphene is given by 
\beq 
\epsilon_g(q,\omega) = 1 - {\tilde V}(q,\omega) \ 
\Pi_0(q,\omega)  \; , 
\label{eps}
\eeq
where ${\tilde V}(q,\omega)$ is the Fourier transform of the screened 
(by the presence of the two slabs) 
Coulomb potential between quasiparticles of graphene 
and $\Pi_0(q,\omega)$ is the first-order dynamical 
polarization of non-interacting 
quasiparticles in graphene. Note that for a very small wavenumber 
$q=\sqrt{q_x^2+q_y^2}$ and a frequency $\omega$ such that 
$v_F q \ll \omega \ll 2\mu/\hbar$ the 
dynamical polarization reads \cite{sols,sarma}
\beq 
\Pi_0(q,\omega) = {\mu\over \pi \hbar^2} {q^2\over \omega^2} \; . 
\label{chi}
\eeq
The collective mode of plasmons in graphene is then obtained from 
the resonance condition \cite{fetter}
\beq 
\epsilon_g(q,\omega) = 0 \; . 
\label{mistic}
\eeq

\section{Perfect conductors}

Let us suppose that the two slabs are perfect conductors. 
A straightforward application of the 
method of image charges \cite{jackson} gives the screened Coulomb potential 
between two particles with electric charge $e$ located 
in the plane $z=0$ at distance $x^2+y^2$ as 
\beq 
V(x,y) = e^2 \sum_{j=-\infty}^{+\infty} 
\Big[ {1\over \sqrt{(x^2+y^2) +(2jL)^2}} 
- {1\over \sqrt{(x^2+y^2)+(2d-2jL)^2}} \Big]\; , 
\eeq
where $x$ and $y$ are Cartesian coordinates in the plane of graphene. 
Performing the Fourier transform we get 
\beq 
{\tilde V}(q) = {2\pi e^2\over q} \sum_{j=-\infty}^{+\infty} 
\left[ \mbox{e}^{-2q|j|L} - \mbox{e}^{-2q|d-jL|} \right] \;  
\eeq
with $q=\sqrt{q_x^2+q_y^2}$. 
The series can be calculated explicity because 
is the sum of geometric series. After straightforward calculations we 
obtain 
\beq 
{\tilde V}(q) = {2\pi e^2\over q} 
\left( 1 - {\mbox{e}^{-2qd} - 2 \mbox{e}^{-2qL} + \mbox{e}^{-2q(L-d)} 
\over 1 - \mbox{e}^{-2qL}}
\right) \; . 
\label{luca}
\eeq

From Eqs. (\ref{eps}), (\ref{chi}) and (\ref{mistic}) the dispersion 
relation of plasmons in graphene can be written as 
\beq 
\omega^2 = {\mu\over \pi \hbar^2 } q^2 {\tilde V}(q) \; ,    
\label{boh}
\eeq
or explicitly 
\beq 
\omega = \sqrt{2\mu e^2\over \hbar^2} \, q^{1/2} \, 
\left( 1 - {\mbox{e}^{-2qd} - 2 \mbox{e}^{-2qL} + \mbox{e}^{-2q(L-d)} 
\over 1 - \mbox{e}^{-2qL}}\right)^{1/2} \; . 
\label{bohboh}
\eeq
Thus we have found an analytical formula for the dispersion 
relation of plasmons in the graphene sheet. 

It is important to observe that, for small $q$, one obtains 
\beq 
{\tilde V}(q) = 4\pi e^2 d \left( 1 - {d\over L}\right) - 
{4\pi e^2\over 3} d^2 L 
\left( 1 - 2{d\over L} + {d^2\over L^2}\right) q^2 + ... 
\label{pio}
\eeq
Consequently, taking into account Eqs. (\ref{boh}) and (\ref{pio}) 
we finally get, for small $q$, the linear dispersion relation 
\beq 
\omega = c_p \ q \; , 
\eeq 
where 
\beq 
c_p= \sqrt{{4\mu e^2 d \over \hbar^2} \left(1-{d\over L}\right)} 
\label{final}
\eeq
is the speed of sound of acustic plasmons in graphene 
sandwiched between two ideal metal gates. Eq. (\ref{final}) is the main 
result of this brief paper. 
The velocity $c_p$ can be controlled by varying the chemical potential 
$\mu$ but also the two distances $d$ and $L$. In the limit $L\to +\infty$ 
from Eq. (\ref{final}) one finds 
\beq 
c_p =  \sqrt{4\mu e^2 d \over \hbar^2}  \;  
\label{start}
\eeq
that is the result of Ref. \cite{principi}, namely the velocity 
of acustic plasmons in graphene coupled to a single ideal metal gate.  

\section{Real materials}

For a generic material the relative dielectric function $\epsilon_m$ 
depends on frequency $\omega$ and wavevector $q$. 
We set $\epsilon_{m,1}(q,\omega)$ 
and $\epsilon_{m,2}(q,\omega)$ the relative dielectric functions of 
lower and upper materials, respectively. In this case the 
derivation of the screened Coulomb potential is slightly 
more complicated but still doable analytically \cite{jackson,new2021}. 
We obtain 
\beqa 
{\tilde V}(q,\omega) = {2\pi e^2\over q} 
\Big( 1 &-& {r_1(q,\omega) \ \mbox{e}^{-2qd} + r_2(q,\omega) \ 
\mbox{e}^{-2q(L-d)} \over 
1 - r_1(q,\omega) \ r_2(q,\omega) \ \mbox{e}^{-2qL}} 
\nonumber 
\\
&+&  {2 \ r_1(q,\omega) \ r_2(q,\omega) \ \mbox{e}^{-2qL} 
\over 1 - r_1(q,\omega) \ r_2(q,\omega) \ \mbox{e}^{-2qL}} 
\Big) \; ,  
\label{flavio}
\eeqa
where 
\beq
r_1(q,\omega) = {\epsilon_{m,1}(q,\omega) -1 \over 
\epsilon_{m,1}(q,\omega) +1} 
\quad\quad
\mbox{ and }
\quad\quad 
r_2(q,\omega) = {\epsilon_{m,2}(q\omega) - 1 \over 
\epsilon_{m,2}(q,\omega) +1}  \; . 
\eeq
Note that for two perfect metal gates, where $r_1=r_2=1$, 
Eq. (\ref{flavio}) becomes exactly Eq. (\ref{luca}). 

\subsection{Materials sticked to graphene}

Setting $L=2d$, in the limit $d\to 0$ Eq. (\ref{flavio}) gives 
\beq 
{\tilde V}(q,\omega) = {2\pi e^2\over \epsilon_{m}(q,\omega) \ q} \; , 
\label{notbad1}
\eeq
where 
\beq 
\epsilon_{m}(q,\omega) = {1\over 2} 
\left( \epsilon_{m,1}(q,\omega) + \epsilon_{m,2}(q,\omega) \right) \; . 
\label{notbad2}
\eeq
Eq. (\ref{notbad1}) is the screened Coulomb potential 
in a graphene sheet between two materials sticked on it. 

We adopt again Eq. (\ref{chi}), which is valid for $q\to 0$ and 
$v_Fq\ll \omega \ll 2\mu/\hbar$ \cite{sols,sarma}, and Eq. (\ref{mistic}). 
Then, for small $q$ and assuming that $\epsilon_m$ is constant, 
we obtain 
\beq 
\omega = \sqrt{2e^2\mu\over \hbar^2 \epsilon_{m}} \ \sqrt{q}  \; , 
\eeq
that is the typical dispersion relation of plasmons in 
graphene exposed to two polar substrates \cite{review}. 

\subsection{Single material slab}

In the absence of the upper slab, i.e. setting $r_2=0$, 
from Eq. (\ref{flavio}) we find 
\beq 
{\tilde V}(q,\omega) = {2\pi e^2\over q} 
\left( 1 - {r(q,\omega) \ \mbox{e}^{-2qd} } \right) 
\eeq
removing the subindex $1$ from $r_1(q,\omega)$. 
Then, for small $q$ and assuming that $r$ is constant, we get 
\beq 
{\tilde V}(q,\omega) = {2\pi e^2(1-r) \over q} 
+ 4\pi e^2 r d  - 
8\pi e^2 r d^2 \ q + ... \; . 
\label{cicredi}
\eeq
Clearly, only if $r=1$ (perfect conductor) the $1/q$ term drops out 
and one again finds acoustic plasmons with speed of sound given 
by Eq. (\ref{start}). More generally, the small-q dispersion relation of 
plasmons reads 
\beq 
\omega = \sqrt{{2\mu e^2 (1-r) \over \hbar^2} q 
+ {4\mu e^2 r d\over \hbar^2} q^2} \; ,  
\eeq
which becomes acoustic-like under the condition 
\beq 
q \gg {(1-r)\over 2d r} \; . 
\eeq 

For a real metal gate the functional dependence 
of $r(q,\omega)$ is crucial. In this case 
the relative dielectric function $\epsilon_m$ can be 
approximated as \cite{mermin}
\beq 
\epsilon_m(q,\omega) = 1 + {q_{TF}^2\over q^2}- 
{\omega_p^2\over \omega^2 + i \Gamma \ \omega} \; , 
\eeq
where $q_{TF}$ is the Thomas-Fermi wavenumber, $\omega_p$ is the plasma 
frequency, and $\Gamma$ the damping constant. Notice that the 
relative dielectric constant of a perfect conductor is 
$\epsilon_m=-\infty$. 

\section{Conclusions}

We have derived a simple formula for the speed of 
sound of the acustic modes of electrons in a graphene sheet. 
The existence of these hydrodynamic effects is due to the presence 
of metallic slabs which induce a screening the Coulomb potential 
of electrons in graphene. Our formula for the graphene 
sandwiched between two metallic slabs generalizes the one obtained 
in Ref. \cite{principi} in the case of graphene coupled to a single 
metallic slab. To conclude, it is important to stress that very recently 
acoustic plasmons have been observed, with a real-space imaging, 
in single graphene sheet over a dielectric-metal slab \cite{new2021}. 
This graphene-dielectric-metal configuration is quite 
different with respect to the one considered in the present paper. 
However, for the sake of completeness, 
in the last section of our paper we have also considered 
the effect of two generic slabs of the screened Coulomb potential of 
two electrons in graphene.

The author acknowledges Marco Polini, Alessandro Principi, 
and Flavio Toigo for useful discussions.

\end{document}